\documentclass[10pt,letterpaper,twocolumn]{article} 

\usepackage{ol2}
\usepackage[draft]{hyperref}
\usepackage{amsmath}

\begin{document}

\twocolumn[ 

\title{Inverse Anderson transition in photonic cages}


\author{Stefano Longhi}
\address{Dipartimento di Fisica, Politecnico di Milano and Istituto di Fotonica e Nanotecnologie del Consiglio Nazionale delle Ricerche, Piazza L. da Vinci 32, I-20133 Milano, Italy (stefano.longhi@polimi.it)}
\address{IFISC (UIB-CSIC), Instituto de Fisica Interdisciplinar y Sistemas Complejos, E-07122 Palma de Mallorca, Spain}

\begin{abstract}
Transport inhibition via Anderson localization is ubiquitous in disordered periodic lattices. However, in crystals displaying only flat bands disorder can lift macroscopic band flattening, removing geometric localization and enabling transport in certain conditions. Such a striking phenomenon, dubbed inverse Anderson transition and predicted for three-dimensional flat band systems, has thus far not been directly observed. Here we suggest a simple quasi one-dimensional photonic flat band  system, namely an Aharonov-Bohm photonic cage, in which correlated binary disorder induces an inverse Anderson transition and ballistic transport.
 \end{abstract}
 ] 

{\em Introduction.} Light propagation in optical waveguide lattices with disorder has provided over the past two decades unique capabilities to realize groundbreaking theoretical concepts of disordered systems, such as Anderson localization and associated phenomena  (see e.g. \cite{r1,r2,r3,r4,r5,r6,r7,r8,r9,r10,r11,r12,r13,r14,r15} and references therein). Anderson localization predicts that static uncorrelated disorder added to a regular lattice can lead to complete localization of wave functions for non-interacting particles and thus the absence of transport. 
Intriguing effects arise when some of the dispersion bands of the clean lattice are flat and the system supports compact localized eigenstates, that are perfectly localized to several lattice sites \cite{r17,r18,r19,r20,r21,r22,r23,r24,r25,r26}. Owing to the diverging effective mass in a flat band lattice, the system becomes very sensitive to disorder and the emerging phenomena can significantly deviate from conventional Anderson localization \cite{r27,r28,r29,r30,r31,r32,r33,r34,r35}. When in the clean lattice a flat band coexists with dispersive bands, weak disorder  hybridizes rather generally the compact flatland states with Bloch waves of dispersive bands, leading to exotic phenomena like localization with unconventional
critical exponents, multifractal behavior and mobility edges with algebraic singularities \cite{
r29,r30,r31}. Remarkably, when \emph{all} the bands of the system are flat, transport in the clean lattice is forbidden by geometric localization and disorder can in principle induce a localization-delocalization transition \cite{r28}. This striking effect, dubbed the inverse Anderson transition, was numerically predicted for a three-dimensional diamond lattice with four-fold degenerated orbitals possessing only flat bands \cite{r28}, however it has thus far not been observed. On the other hand, in low-dimensional systems with entire flat bands geometric localization seems to be robust against uncorrelated static disorder \cite{r35}, preventing the observation of an inverse Anderson transition.\\ 
In this Letter we suggest a rather simple quasi one-dimensional (1D) photonic system displaying flat bands in the clean limit, namely a photonic analogue of the Aharonov-Bohm cage \cite{r19,referee1,referee2,r36,r37}, and show analytically that \emph{correlated binary disorder} enables ballistic transport and absolutely continuous spectrum with dispersive bands, thus providing an experimentally simple and accessible system to observe inverse Anderson localization in a low-dimensional system.\\
\\
{\em Aharonov-Bohm photonic cage with disorder.} 
We consider a photonic cage \cite{r19,r36,r37} consisting of a quasi 1D rhombic lattice of evanescently-coupled optical waveguides with three sublattices A, B and C, in which a synthetic magnetic flux $\varphi$ is applied in each closed square loop via Peierls' substitution of the coupling constant between waveguides of sublattices A and B, as shown in Fig.1(a). 
We assume static on-site potential disorder $V_n$ and $W_n$ in the outer sublattices B and C, which corresponds to disorder in the propagation constant shift of waveguide modes in sublattices B and C with respect to waveguides in sublattice A.
In the nearest-neighbor tight-binding approximation, light propagation in the rhombic lattice is described by the following coupled-mode equations for the modal amplitudes $a_n$, $b_n$ and $c_n$ in the various guides
\begin{eqnarray}
i \frac{da_n}{dz} & = &  \kappa \left( b_n \exp(i \varphi)+b_{n-1}+c_{n}+c_{n-1} \right)  \nonumber \\
i \frac{db_n}{dz} & = & \kappa \left( a_n \exp(-i \varphi) + a_{n+1} \right)+V_n b_n  \\
i \frac{dc_n}{dz} & = & \kappa \left( a_n + a_{n+1} \right)+W_n c_n \nonumber
\end{eqnarray}
 where $\kappa$ is the coupling constant between adjacent waveguides, $z$ is the propagation (axial) distance, and $\varphi$ is the synthetic magnetic flux. The clean lattice ($V_n=W_n=0$) sustains three bands with the dispersion relations given by 
\begin{equation}
E_0=0 \; , \; \; \; E_{\pm}= \pm 2 \kappa \sqrt{1+\cos (\varphi  /2) \cos(q+\varphi /2)}
\end{equation}
where $-\pi \leq q < \pi$ is the Bloch wave number. For $\varphi= \pi$, the spectrum is made up of three 
{\it flat} bands, $E_0=0$ and $E_{\pm}= \pm 2 \kappa$. The minimally extended (compact)
eigenstates corresponding to the three flat bands are displayed in Fig.1(b). In this case
a complete suppression of any wave packet spreading in the lattice is thus realized, corresponding to the so-called Aharonov-Bhom cage. The required $\varphi= \pi$ synthetic magnetic flux can be experimentally realized either by suitable bending engineering of the lattice \cite{r19,r36} or by using an auxiliary lattice to indirectly couple waveguides in sublattices A and B \cite{r37}.\\
In the presence of on-site potential disorder in the outer sublattices B and C, after setting $(a_n,b_n,c_n)^T=(A_n,B_n,C_n)^T \exp(-i Ez)$ in Eq.(1) and eliminating the variables $B_n$ and $C_n$ in the equations so obtained, one can write the following eigenvalue equation for the energy spectrum $E$
\begin{equation}
E A_n= \Delta_n(E) A_n+ J_n(E) A_{n+1}+J_{n-1}^*(E) A_{n-1} 
\end{equation}
where we have set 
\begin{eqnarray}
\Delta_n(E) & = & \frac{\kappa^2}{E-W_n}+\frac{\kappa^2}{E-W_{n-1}}+\frac{\kappa^2}{E-V_{n}}+ \nonumber \\
& + &  \frac{\kappa^2}{E-V_{n-1}}\\
J_n (E) & = & \frac{\kappa^2}{E-W_n}+ \frac{\kappa^2 \exp(i \varphi)}{E-V_n}.
\end{eqnarray}
 \begin{figure}[htb]
 \centerline{\includegraphics[width=8.7cm]{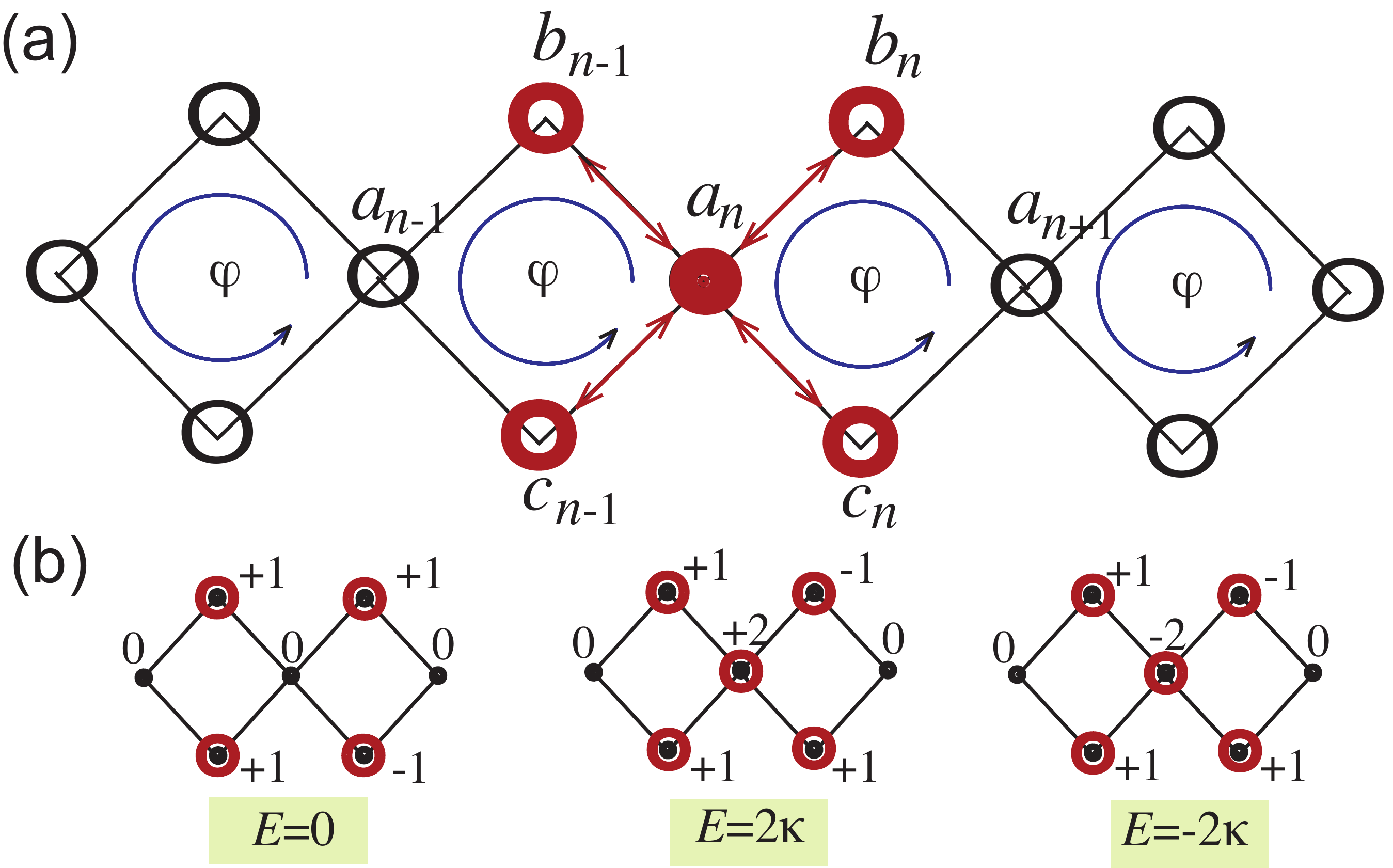}} \caption{ \small
 (Color online) (a) Schematic of a quasi 1D photonic cage made of  of three sublattices A, B and C of optical waveguides arranged in a rhombic geometry. A  synthetic magnetic flux $\varphi$ is applied in each plaquette. At $\varphi= \pi$ the lattice sustains the three flat bands $E_0=0$, $E_{\pm}= \pm 2 \kappa$, and Aharonov-Bohm caging is realized: a photon initially localized on the central guide (filled circle) oscillates to its nearest neighbors (empty circles), but will not propagate through the lattice. (b) Compact localized eigenstates of energies $E_0=0$ and $E_{\pm}= \pm 2 \kappa$ for $\varphi= \pi$.}
 \end{figure} 
Equation (3)  formally describes the spectral problem of a 1D tight-binding lattice with nearest-neighbor hopping with an energy-dependent disorder in both on-site potential $\Delta_n(E)$ and hopping amplitudes $J_n(E)$. The impact of uncorrelated disorder with a continuous probability density function in the
uniform rhombic lattice of Fig.1(a), with and without the synthetic magnetic flux $\varphi$, was investigated in some previous works \cite{r27,r32,r35}. For $\varphi \neq \pi$, i.e. when the system sustains one flat band ($E_0=0$) and two dispersive bands ($E_{\pm}$), static disorder rather generally removes the flat band eigenvalue degeneracy and provides mixing inside this band, as well as with states originating from the other dispersive bands \cite{r31,r32}. 
In this work we focus our attention to the flat band case $\varphi= \pi$, since we wish to establish whether \emph{static} disorder can induce transport in the lattice (inverse Anderson transition \cite{r28}). For $\varphi= \pi$, a recent study \cite{r35} based on extended numerical results indicates that static on-site uncorrelated disorder can not induce transport,  and that two different localization mechanisms, namely frustration (geometric) and Anderson (exponential) localization, do compete: For weak static on-site disorder (smaller than $\sim 2 \kappa$), the eigenstates from each band are separated by gaps and their localization lengths saturate and do not depend on the disorder
strength, indicating that geometric localization prevails over Anderson localization.  On the other hand, for strong on-site disorder  (larger than $ \sim 2 \kappa$) the energy bands are mixed and the localization length decreases as the disorder strength is increased (like in ordinary Anderson localization).\\
\\
{\em Transport in the photonic cage with antisymmetric correlated disorder.}
 An open question is whether other kinds of on-site static disorder can induce transport. Inspired by the random dimer model \cite{r38}, we assume \emph{correlated} disorder in sublattices B and C, by considering either \emph{symmetric} ($W_n=V_n$) or \emph{antisymmetric} ($W_n=-V_n$) correlated disorder, where $V_n$ are independent stochastic variables with the same probability density function $f(V)$ of zero mean. As shown in the Supplemental document, in the former case the disorder lifts band degeneracy, however compact localized states are robust and transport is thus prevented. A more interesting scenario arises in the antisymmetric correlated disorder $W_n=-V_n$. In this case,  provided that $V_n \neq 0$ from Eq.(4) it follows that $\Delta_n(E)$ vanishes at zero energy $E=0$, i.e. $\Delta_n(E=0)=0$, and Eq.(3) at $E=0$ reads
\begin{equation}
\frac{1}{V_n} A_{n+1}+\frac{1}{V_{n-1}}A_{n-1}=0
\end{equation}
from which the right/left Lyapunov exponents 
\begin{equation}
\mu_{\pm}(E=0)= \lim_{n \rightarrow \pm \infty} \frac{1}{n} \log \left|  \frac{A_n}{A_{0,1}} \right| 
\end{equation}
can be readily computed, yielding $\mu_{\pm}(E=0)=0$. This means that $E=0$ belongs to the energy spectrum and the two corresponding linearly-independent wave functions, recursively defined by Eq.(6) assuming either $A_0=1$ and $A_{2n+1}=0$ or $A_1=1$ and $A_{2n}=0$, are \emph{extended} states. We stress that such a result holds provided that $V_n \neq 0$, i.e. for any probability density function $f(V)$ vanishing at $V=0$. The existence of extended states at the zero energy indicates that disorder-induced transport in the photonic cage system is possible. We checked this prediction by considering light transport in the photonic cage structure with antisymmetric correlated disorder described by the probability density function 
\begin{equation}
f(V)=	\left\{
\begin{array}{ll}
\frac{1}{ 2 \Delta} & |V \pm \mathcal{V}|< \Delta /2 \\
0 & {\rm otherwise}
\end{array}
\right. ,
\end{equation}
with $\Delta < 2 \mathcal{V}$. Note that, in the limit $\Delta \rightarrow 2 \mathcal{V}$, $f(V)$ is uniformly distributed in the range $(-2 \mathcal{V}, 2 \mathcal{V})$, and 
we expect localization and the absence of transport in this limit \cite{r35}.  In the other limit $\Delta \rightarrow 0$, $f(V)$ is the Bernoulli distribution since $V_n$ can take only the two values $\pm \mathcal{V}$ with the same probability.
As an example, in Figs.2(a) and (b) we show the numerically-computed light propagation dynamics in the waveguide photonic cage system
with initial condition corresponding to the excitation of the waveguide of sublattice A at site $n=0$, i.e. $a_n(0)=\delta_{n,0}$ and $b_n(0)=c_n(0)=0$. The spreading of the discretized light beam, shown in Fig.2(b), is measured by the variance

\[ \sigma^2(z)= \sum_n n^2 (|a_n|^2+|b_n|^2+|c_n|^2), \]
 with $\sigma(z) \sim z$ for ballistic transport. To characterize the localization properties of the wave functions, we consider the inverse participation ratio (IPR), which can  distinguish localized and extended states. Assuming a finite lattice with $N$ unit cells, for a normalized wave function the IPR is defined as 
\begin{equation}
\text{IPR} =\sum_{n=1}^{N}   \left( \left| A_{n}\right|^{4} + \left| B_{n}\right|^{4} +\left| C_{n}\right|^{4} \right).
\end{equation}
The IPR of an extended state scales as $ \sim 1/N$, hence vanishing in the large $N$ limit, while it remains finite for a localized state. Figure 2(c) shows the numerically-computed energy spectrum and IPR in a lattice comprising $N=300$ unit cells assuming periodic boundary conditions. Clearly, disorder lifts the degeneracy of the three energy levels (flat bands) $E=0, \pm 2 \kappa$ of the clean system, with clusters of energies separated by two main gaps and with small IPR of wave functions far from the band edges, indicating that there are many extended states (besides the ones at energy $E=0$).\\ 
 \begin{figure}[htb]
 \centerline{\includegraphics[width=8.7cm]{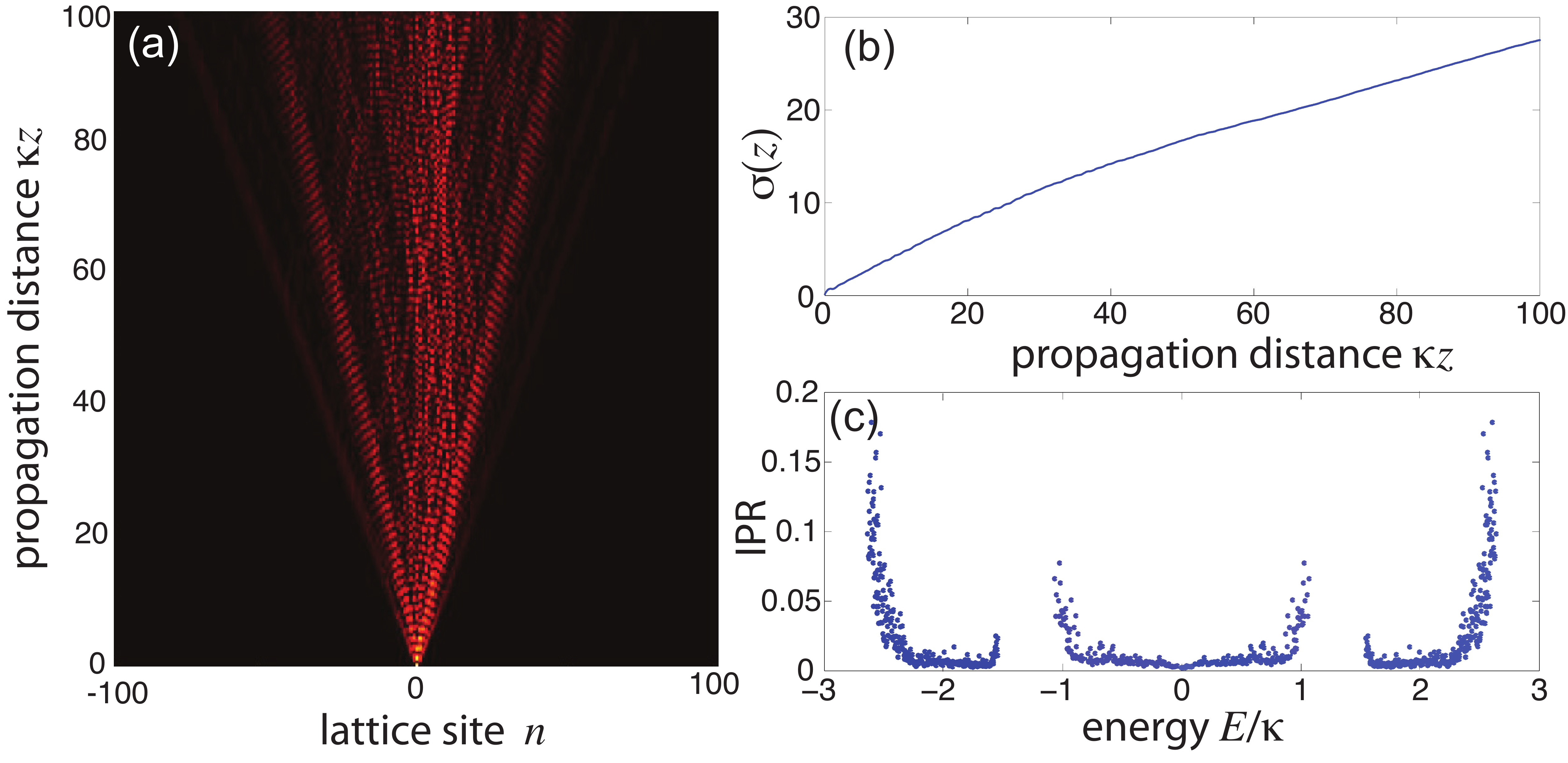}} \caption{ \small
 (Color online) (a,b)  Light propagation in a photonic cage lattice with correlated antisymmetric disorder $W_n=-V_n$. The density probability function $f(V)$ of the stochastic variable $V_n$ is given by Eq.(8) with $\mathcal{V}=\kappa$ and $\Delta=\kappa/2$. The lattice is initially excited in the waveguide of sublattice A at site $n=0$. Panel (a) shows the numerically-computed evolution of modal amplitudes $|a_n(z)|$ versus normalized propagation distance $ \kappa z$ on a pseudo color map for a given realization of disorder. The corresponding evolution of the standard deviation $\sigma(z)$ is depicted in panel (b). Panel (c) shows the numerically-computed behavior of the IPR for the eigenstates of Eq.(1) versus normalized energy $E/ \kappa$ in a lattice with $N=300$ unit cells under periodic boundary conditions.}
 \end{figure}
The spreading in the lattice is controlled by the ratio $\Delta / \mathcal{V}$, as shown in Fig.3. The figure depicts the numerically-computed evolution of the standard deviation $\sigma(z)$, averaged over 50 different realizations of disorder, for a few decreasing values of $\Delta / \mathcal{V}$. The right panels in Fig.3 show typical light dynamics evolution in the lattice. Localization is found at $\Delta=2 \mathcal{V}$ (uniform distribution), while the fastest spreading is observed when $\Delta=0$ (Bernoulli distribution).  The Bernoulli distribution ($\Delta \rightarrow 0$) is particularly interesting since it is exactly solvable, proving in a rigorous way that inverse Anderson localization with ballistic transport arises in the photonic cage structure with correlated antisymmetric disorder. In fact, for a Bernoulli distribution $V_n$ can take only the two values $\pm \mathcal{V}$ with probabilities $p$ and $q=1-p$. In this case, from Eq.(4) it follows that $\Delta_n(E)=4 \kappa^2 E/(E^2-\mathcal{V}^2)$ is independent of site index $n$, whereas form Eq.(5) one has $J_n=-2 \kappa^2 \mathcal{V} \exp(i \pi \delta_n) /(E^2-\mathcal{V}^2)$, with $\delta_n=0, 1$ for $V_n= \pm \mathcal{V}$. 
After introduction of the gauge transformation 
\begin{equation}
A_n= \bar{A}_n \exp \left( -i \pi \sum_{l=0}^{n-1} \delta_l \right)
\end{equation}
for the amplitudes $A_n$, in the new variables $\bar{A}_n$ the eigenvalue equation (3) takes the form
\begin{figure}[htb]
 \centerline{\includegraphics[width=8.7cm]{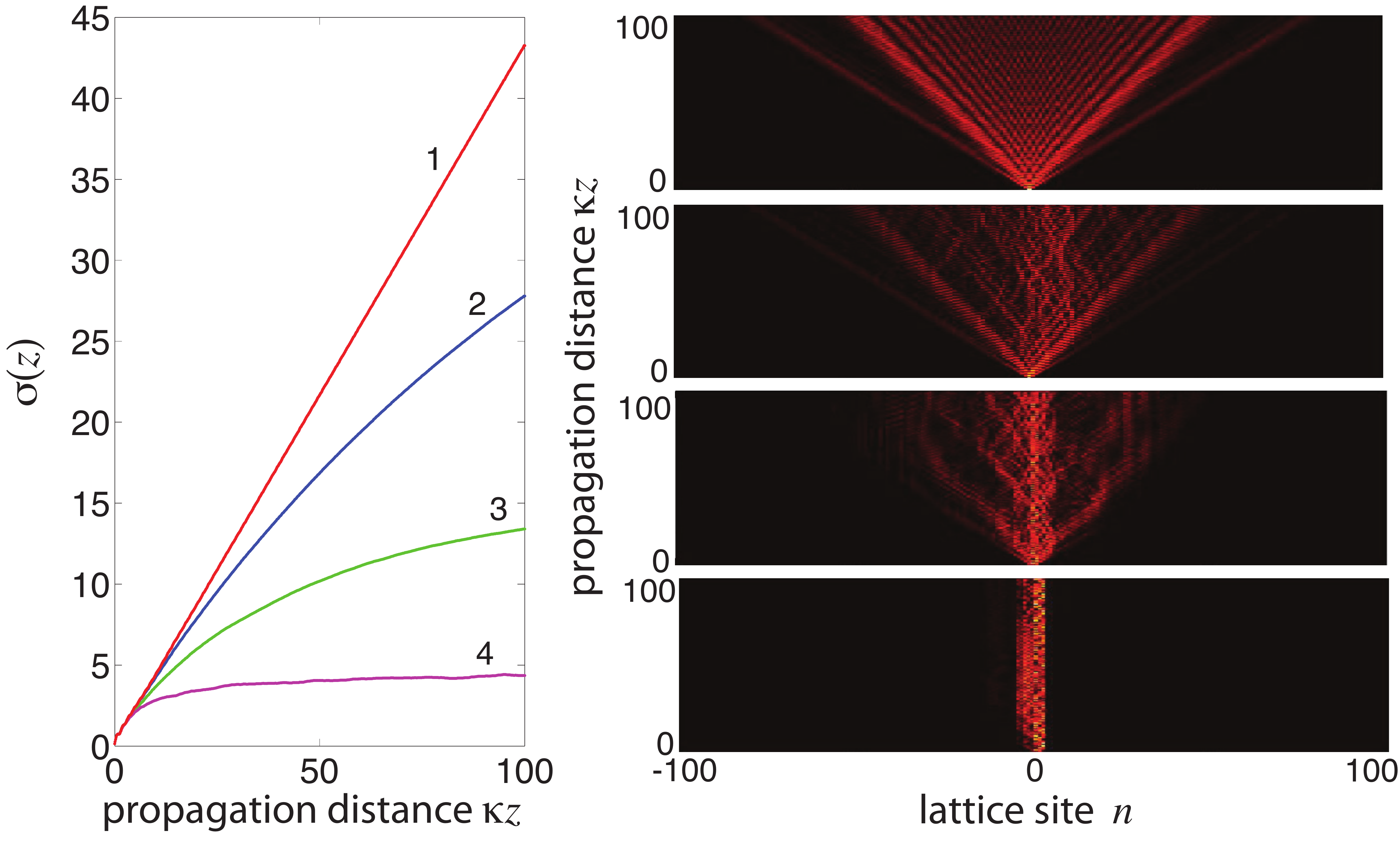}} \caption{ \small
 (Color online) Numerically-computed evolution of the standard deviation $\sigma(z)$  for light propagation in a disordered photonic cage lattice, averaged over 50 different realizations of disorder. The probability density function is given by Eq.(8) with $\mathcal{V}=\kappa$, while a few increasing values of $\Delta$ are considered. Curve 1: $\Delta=0$ (Bernoulli distribution); curve 2: $\Delta= \mathcal{V} /2$; curve 3: $\Delta=\mathcal{V}$; curve 4: $\Delta=2 \mathcal{V}$ (uniform distribution). The right panels show typical light dynamics in the lattice on a pseudo color map (increasing values of $\Delta$ from top to bottom).}
 \end{figure}

\begin{equation}
E \bar{A}_n= \frac{4 \kappa^2 E}{E^2-\mathcal{V}^2} \bar{A}_n-\frac{2 \kappa^2 \mathcal{V}}{E^2-V^2} \left( \bar{A}_{n+1}+ \bar{A}_{n-1}  \right)
\end{equation}
with energy-dependent \emph{but site-independent} hopping amplitudes and on-site potential. The wave functions in the amplitudes $\bar{A}_n$ are thus extended and of Bloch type, $\bar{A}_n=\bar{A} \exp(iq n)$, with energy dispersion $E(q)$ satisfying the condition
\begin{equation}
E=\frac{4 \kappa^2}{E^2-\mathcal{V}^2} \left( E- \mathcal{V} \cos q \right)
\end{equation}
 i.e. the cubic equation
 \begin{equation}
 E^3-(\mathcal{V}^2+4 \kappa^2)E+ 4 \kappa^2 \mathcal{V} \cos q =0.
 \end{equation}
 Therefore, for correlated binary (Bernoulli) antisymmetric disorder the energy spectrum remains absolutely continuous and composed by three \emph{dispersive} Bloch bands, with energy dispersion defined by the roots of Eq.(13). The width of the  spectrum $\Delta E$, defined as the sums of the widths of the three dispersive Bloch bands, vanishes for $\mathcal{V}=0$, i.e. in the clean limit corresponding to the flat bands, and  in the strong disorder limit $\mathcal{V} \gg \kappa$, as shown in Fig.4(a). The largest bandwidth $\Delta E$ is attained at $\mathcal{V}= \sqrt{2} \kappa$, at which the two gaps separating the three bands vanish. At this amplitude of disorder, the transport is fastest, as shown in Fig.4(b). As a final comment, we note that, if on-site potential disorder $Y_n$ were considered also in sublattice A, a term $Y_n$ should be added to the effective on-site potential $\Delta_n(E)$, leading to localization and thus preventing transport. 
  \begin{figure}[htb]
 \centerline{\includegraphics[width=8.7cm]{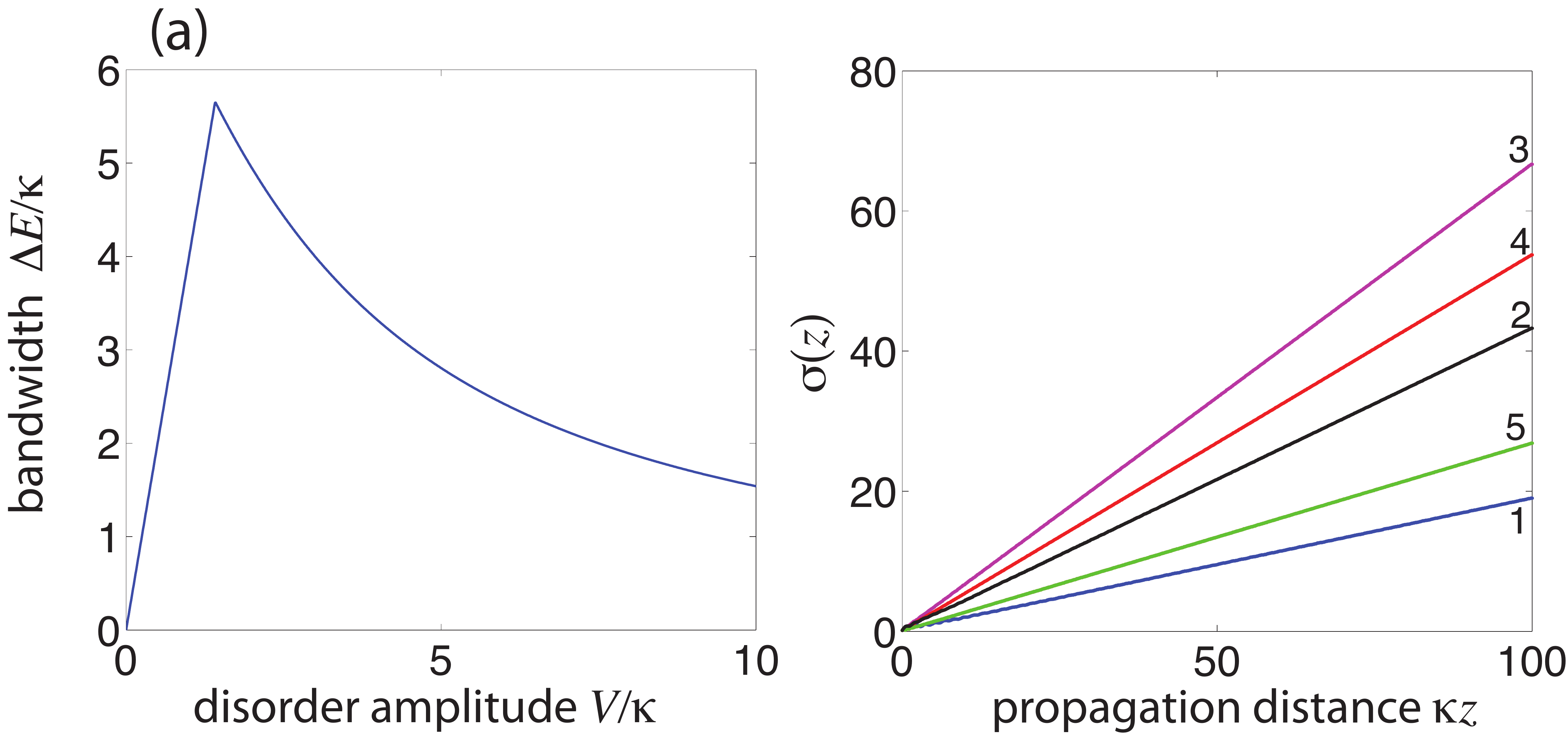}} \caption{ \small
 (Color online) (a) Behavior of the bandwidth $\Delta E$ of the dispersive bands versus disorder amplitude $\mathcal{V}$ for a binary (Bernoulli) distribution of $V_n$. (b) Behavior of the standard deviation $\sigma(z)$ versus propagation distance $z$ for a few increasing values of $\mathcal{V}$. Curve 1: $\mathcal{V} / \kappa=0.5$; curve 2: $\mathcal{V} / \kappa=1$; curve 3: $ \mathcal{V} / \kappa= \sqrt{2}$; curve 4: $ \mathcal{V} / \kappa=4$; curve 5: $ \mathcal{V}/ \kappa=10$. The fastest wave spreading is found at $\mathcal{V} / \kappa= \sqrt{2}$.}
 \end{figure}

The inverse Anderson transition in a photonic cage predicted in this work should be feasible for an experimental observation using optical waveguide lattices realized by the femtosecond laser writing technology in bulk glasses \cite{r36,r37}. In the Supplemental document we provide results on wave spreading for realistic parameter values.

{\em Conclusions.} Inverse Anderson transition arises in certain lattices displaying fully flat bands, where disorder removes geometric localization and restores transport. In this work we suggested a quasi 1D photonic system where inverse Anderson transition could be observed. Such a system is interesting because: (i) Inverse Anderson transition can be demonstrated in an exact (analytical) way; (ii) It highlghts the role of correlated disorder; (iii) It is feasible for an experimental demonstration using photonic waveguide lattices, paving the way toward the first experimental observation of the inverse Anderson transition.
The present results unravel new insights on the interplay between correlated disorder and flat band systems, and could be of interest beyond photonics.\\


\begin{thebibliography}{99}



\bibitem{r1}
T. Schwartz, G. Bartal, S. Fishman, and M. Segev, Nature {\bf 446}, 52 (2007).

\bibitem{r2}
Y. Lahini, A. Avidan, F. Pozzi, M. Sorel, R. Morandotti, D.N. Christodoulides, and Y. Silberberg, Phys. Rev. Lett. {\bf 100}, 013906 (2008).

\bibitem{r3}
 L. Martin, G. Di Giuseppe, A. Perez-Leija, R. Keil, F. Dreisow, M. Heinrich, S. Nolte, A. Szameit, A.F. Abouraddy, D.N. Christodoulides, and B.E.A. Saleh,
 Opt. Express {\bf 19}, 13636 (2011).
 
 \bibitem{r4}
D.M. Jovic, Y.S. Kivshar, C. Denz, and M.R. Belic, Phys. Rev. A {\bf 83}, 033813 (2011).
 
 \bibitem{r5}
D. Jovic, C. Denz, and M. Belic, Opt. Photon. News {\bf 22}, 34 (2011).
 
\bibitem{r6} 
U. Naether, Y.V. Kartashov, V.A. Vysloukh, S. Nolte, A. T\"unnermann, L. Torner, and A. Szameit, Opt. Lett. {\bf 37}, 593 (2012).

\bibitem{r7}
S. St\"utzer, Y. V. Kartashov, V. A. Vysloukh, A. T\"unnermann, S. Nolte, M. Lewenstein, L. Torner, and A. Szameit, Opt. Lett. {\bf 37}, 1715 (2012).

\bibitem{r8}
M. Segev, Y. Silberberg, and D.N. Christodoulides, Nature Photon. {\bf 7}, 197 (2013).

\bibitem{r9}
P. Titum, N.H. Lindner, M.C. Rechtsman, and G. Refael,  Phys. Rev. Lett. {\bf 114}, 056801 (2015).

\bibitem{r10}
F. Baboux, L. Ge, T. Jacqmin, M. Biondi, E. Galopin, A. Lemaitre, L. Le Gratiet, I. Sagnes, S. Schmidt, H.E. T\"ureci, A. Amo, and J. Bloch, Phys. Rev. Lett. {\bf 116}, 066402 (2016).

\bibitem{r11}
I. D. Vatnik, A. Tikan, G. Onishchukov, D.V. Churkin, and A.A. Sukhorukov,  Sci. Rep. {\bf 7}, 4301 (2017).

\bibitem{r12}
S. St\"utzer, Y. Plotnik, Y. Lumer, P. Titum, N. H. Lindner, M. Segev, M. C. Rechtsman, and A. Szameit, Nature {\bf 560}, 461 (2018).

\bibitem{r13}
 P. Wang, Y. Zheng, X. Chen, C. Huang, Y.V. Kartashov, L. Torner, V.V. Konotop, and F. Ye, 
 Nature {\bf 577}, 42 (2020).

\bibitem{r14}
S. Longhi, Opt. Lett. {\bf 45}, 4036 (2020).

\bibitem{r15}
 D. Guzman-Silva, M. Heinrich, T. Biesenthal, Y. V. Kartashov, and A. Szameit, Opt. Lett. {\bf 45}, 415 (2020).
 
  
 \bibitem{r17} 
 D. Leykam and S. Flach,  APL Photon. {\bf 3}, 070901 (2018).
 \bibitem{r18}
 L. Tang , D. Song , S. Xia , S. Xia , J. Ma , W. Yan , Y. Hu , J. Xu , D. Leykam, and Z. Chen,  Nanophoton. {\bf 9}, 1161 (2020).
 \bibitem{r19}
S. Longhi,  Opt. Lett. {\bf 39}, 5892 (2014).
\bibitem{r20}
R. A. Vicencio, C. Cantillano, L. Morales-Inostroza, B. Real, C. Meijia-Cortes, S.Weimann, A. Szameit, and M. I. Molina,
Phys. Rev. Lett. {\bf 114}, 245503 (2015).
\bibitem{r21}
S. Mukherjee, A. Spracklen, D. Choudhury, N. Goldman, P. Ohberg, E. Andersson, and R. R. Thomson, Phys. Rev. Lett. {\bf 114}, 245504 (2015).
\bibitem{r22}
S. Xia, Y. Hu, D. Song, Y. Zong, L. Tang, and Z. Chen,  Opt. Lett. {\bf 41}, 1435 (2016).
\bibitem{r23}
S. Mukherjee and R. R. Thomson, Opt. Lett. {\bf 40}, 5443 (2015).
\bibitem{r24}
E. Travkin, F. Diebel, and C. Denz, Appl. Phys. Lett. {\bf 111}, 011104 (2017).
\bibitem{r25}
B. Real, C. Cantillano, D. Lopez-Gonzalez, A. Szameit, M. Aono, M. Naruse, S.-J. Kim, K. Wang, and R. A. Vicencio,
Sci. Rep. {\bf 7}, 15085 (2017).
\bibitem{r26}
S. Mukherjee and R. R. Thomson, Opt. Lett. {\bf 42}, 2243 (2017).
\bibitem{r27}
J. Vidal, P. Butaud, B. Doucot, and R. Mosseri, 
Phys. Rev. B {\bf 64}, 155306 (2001).
\bibitem{r28}
M. Goda, S. Nishino, and H. Matsuda, 
Phys. Rev. Lett. {\bf 96}, 126401 (2006).
\bibitem{r29}
J. T. Chalker, T. S. Pickles, and P. Shukla, Phys. Rev. B {\bf 82},
104209 (2010).
\bibitem{r30}
S. Flach, D. Leykam, J. D. Bodyfelt, P. Matthies, and A.
S. Desyatnikov, EPL {\bf 105}, 30001 (2014).
\bibitem{r31}
J.D. Bodyfelt, D. Leykam, C. Danieli, X. Yu, and S. Flach,  Phys. Rev. Lett. {\bf 113}, 236403 (2014).
\bibitem{r32}
D. Leykam, J.D. Bodyfelt, A.S. Desyatnikov, and S. Flach,  Eur. Phys. J. B {\bf 90}, 1 (2017).
\bibitem{r33}
P. Shukla, Phys. Rev. B {\bf 98}, 054206 (2018).
\bibitem{r34}
C. Gneiting, Z. Li, and F. Nori, Phys. Rev. B {\bf 98}, 134203 (2018).
\bibitem{r35}
G. Gligoric, D. Leykam, and A. Maluckov, 
Phys. Rev. A {\bf 101}, 023839 (2020).
\bibitem{referee1}
J. Vidal, R. Mosseri, and B. Doucot, Phys. Rev. Lett. {\bf 81}, 5888 (1998).
\bibitem{referee2}
J. Vidal, B. Doucot, R. Mosseri, and P. Butaud,
Phys. Rev. Lett. {\bf 85}, 3906  (2000).
\bibitem{r36}
S. Mukherjee, M. Di Liberto, P. Ohberg, R.R. Thomson, and N. Goldman, Phys. Rev. Lett. {\bf 121}, 075502 (2018).
\bibitem{r37}
M. Kremer, I. Petrides, E. Meyer, M. Heinrich, O. Zilberberg, and A. Szameit, Nature Commun. {\bf 11}, 907 (2020).
\bibitem{r38}
D.H. Dunlap, H.-L. Wu, and P. Phillips, Phys. Rev. Lett. {\bf 65}, 88 (1998).



\end{thebibliography}
\end{document}